\documentclass[12pt]{article}
\usepackage{epsf,epsfig}
\usepackage{color}
\usepackage{amssymb}
\usepackage{subfigure}

\def\la{~\mbox{\raisebox{-.6ex}{$\stackrel{<}{\sim}$}}~}
\def\ga{~\mbox{\raisebox{-.6ex}{$\stackrel{>}{\sim}$}}~}

\begin{document}

\begin{titlepage}
\vfill
\begin{flushright}
\today
\end{flushright}

\vfill
\begin{center}
\baselineskip=16pt
%
{\Large\bf Dynamics of localized Kaluza-Klein black holes}
\vskip 0.15in
{\Large\bf  in a collapsing universe}
\vskip 0.5cm
{\large {\sl }}
\vskip 10.mm
{\bf 
David Kastor, Lorenzo Sorbo and Jennie Traschen\\

\vskip 1cm
{
	Department of Physics, University of Massachusetts, Amherst, MA 01003\\	
	\texttt{kastor@physics.umass.edu, sorbo@physics.umass.edu, traschen@physics.umass.edu}
     }}
\vspace{6pt}
\end{center}
\vskip 0.2in
\par
\begin{center}
{\bf Abstract}
 \end{center}
The Clayton Antitrust Act of 1914 prohibits corporate mergers that would result in certain highly undesired end states.
We study an exact solution of the Einstein equations describing localized, charged Kaluza-Klein black holes in a collapsing deSitter universe and seek to demonstrate that a similar effect holds, preventing  a potentially catastrophic  black hole merger.   As the collapse proceeds, it is natural to expect that the black hole undergoes a topological transition, wrapping around the shrinking compact dimension to merge with itself and form a black string.  However, the putative uniform charged black string end state is singular and such a transition would violate (a reasonable notion of) cosmic censorship.  We present analytic and numerical evidence that strongly suggests the absence of such a transition.   Based on this evidence, we expect that the Kaluza-Klein black hole horizon stays localized, despite the increasingly constraining size of the compact dimension.  On the other hand, the deSitter horizon does change between spherical and cylindrical topologies in a simple way.
 \begin{quote}
  \vfill
\vskip 2.mm
\end{quote}
\end{titlepage}



\section{Introduction}

The Clayton Antitrust Act, enacted by the U.S. Congress in 1914, prohibits corporate mergers that would be deleterious to the public good, {\it e.g. } by substantially reducing competition in a given industry and/or bringing about monopolistic pricing power.  In this paper, we will discuss a system in general relativity in which the invisible hand of the cosmic censor appears to similarly prevent undesireable mergers, in this case mergers of black holes that would lead to a singular geometry.

The mergers we will discuss are actually self-mergers.  We consider black holes in five spacetime dimensions with Kaluza-Klein boundary conditions, {\it i.e.} with one spatial dimension compactified.  This system has been extensively studied (see \cite{Kol:2004ww,Harmark:2005pp,Horowitz:2011cq} for reviews).
Black holes can either wrap the compact direction and have horizon topology $S^2\times S^1$ or they can be localized in the compact direction with horizon topology $S^3$.   

If we imagine fixing the mass $M$ of the black hole, then for sufficiently large size $L$ of the compact dimension a localized black hole should be well approximated by the five dimensional Schwarzschild metric.  
If $L$ is decreased, then the horizon of the localized black hole will gradually distort, becoming elongated in the compact direction, in response to its own gravitational pull.  Eventually, the black hole will merge with itself around the compact direction and for sufficiently small $L$ localized black holes with the fixed mass $M$ will not exist (again see the review \cite{Kol:2004ww,Harmark:2005pp,Horowitz:2011cq} and also the original numerical work in \cite{Kudoh:2003ki,Kudoh:2004hs}).

This has been a description of the behavior of a continuous family of static solutions to the equations of motion.  It would also be fascinating to be able to study this merger in a dynamical setting, with a compact direction that actually shrinks with time.  Remarkably, an analytic solution describing such localized black holes in a shrinking compact direction can be constructed, if we include two additional ingredients.  We must consider charged black holes, which allows for the construction of localized black holes exactly, and also a positive cosmological constant $\Lambda$, which provides the dynamics.  

The starting point for our work is the deSitter multi-black hole solution of \cite{Kastor:1992nn} and its generalization to five dimensions given in  \cite{London:1995ib}.  The black holes can be arranged into a periodic one-dimensional array and then identified to form a compact dimension.
We then consider the evolution of this localized black hole in the collapsing phase of the resulting Kaluza-Klein-deSitter (KKdS) background.  We will additionally always work in the limit  where the parameter $\epsilon\equiv MH_0^2$, where $H_0=\sqrt{\Lambda/6}$ is the Hubble constant, is much less than one.  In this limit, with the black hole horizon size much smaller than the Hubble scale, it is known \cite{Brill:1993tm} that a pair of initially separate black holes as in \cite{Kastor:1992nn}  will merge into a larger black hole.  This merger process was studied in detail numerically in \cite{Nakao:1994mm} at the level of apparent horizons.

We shall also focus on apparent horizons in our study of KKdS black hole spacetimes.   Our results, however, initially seem to be different than expected.  In the early time limit, the compact dimension is very large and the black hole horizon is a sphere of small coordinate radius.   As the scale factor $a(t)$ decreases, the coordinate size of the horizon grows for a time and then hangs up at a finite value smaller than the compactification length $L$, with the overall area of the horizon remaining approximately fixed.  We find no evidence that the black hole horizon self-merges into one that wraps the compact direction.

In retrospect, this is not so surprising a result.  In considering whether our charged, localized KKdS black hole should evolve to a final state with a horizon wrapping the compact direction, we should ask whether such a well-behaved solution with a cylindrical horizon exists.  In the case of a pair of black holes satisfying $\epsilon_i= M_iH_0^2\ll 1$ for $i=1,2$ considered in \cite{Brill:1993tm,Nakao:1994mm}, the initially separate horizons simply merge into that of a larger black hole with mass $M=M_1 +M_2$.   The coordinate radii of the initially separate apparent horizons, in this case, continue expanding with time as the universe collapses and eventually merge as 
the metric approaches that of the black hole metric with mass $M$.  The localized KKdS black hole metric does in the far field limit approach that of a uniform charged black string wrapping the compact direction, but differs in the near horizon regime.  As we discuss below, the horizon of this black string is singular.
Our results strongly indicate that the KKdS black hole horizon remains regular and localized and does not self-merge into this end state.   The cosmic censor apparently  acts  to prevent such a singular result.  On the other hand, we will see that the topology of the deSitter horizon does evolve from spherical to cylindrical  in a particularly simple way.

The paper will proceed in the following way.  In section~\ref{bhsection} we present the charged localized KKdS black hole spacetimes and discuss some of their basic properties.  In section~\ref{apparent} we begin the discussion of apparent horizons by studying their behavior in certain limiting cases, including the KKdS spacetime with no black holes and the single RNdS black hole spacetime (which can be thought of as the limit of the KKdS black hole when the compactification length is taken to infinity).  We are concerned with both black hole and cosmological horizons.  The intuition we build in this section informs our analysis of the apparent horizons of the KKdS black hole in section~\ref{kkdshorizons}.  In section~\ref{string} we turn to the analysis of the uniform charged KKdS black string, including our argument that it has a singular horizon.

\section{Kaluza-Klein-DeSitter black holes}\label{bhsection}

The KKdS black hole spacetimes we will study are built by taking a one dimensional array of charge equal mass RNdS black holes in D=5  \cite{Kastor:1992nn,London:1995ib}.  Here we present these metrics and a number of their basic properties.
 We consider $D=5$ Einstein-Maxwell theory  with positive cosmological constant $\Lambda$, described by the action 
\begin{equation}\label{action}
S=\int d^5 x \sqrt{-g} \left(\frac{1}{2}\,R-\Lambda-\frac{3}{4}\,F^2\right).
\end{equation}
For the corresponding $D=4$ system, Kastor and Traschen (KT) have found a metric describing an arbitrary number of charged black holes. The generalization of the KT metrics \cite{Kastor:1992nn}  to arbitrary higher dimensions was found in \cite{London:1995ib} and the $D=5$ case is given by
\begin{equation}\label{basicmetric}
ds^2 = -{1\over U^2} dt^2 + a^2(t)\, U\,d\vec x\cdot d\vec x, \qquad A=\pm {1 \over U} dt
\end{equation}
where $a(t)=\exp(Ht)$ with $H=\pm H_0$ and $H_0^2=\Lambda/6$, and the metric function $U(\vec x,t) $ has the multi-center form
\begin{equation}
U(\vec x,t) = 1 +\sum_{i=1}^N {M_i\over a^2(t)\, r_i^{2}},\qquad r_i^2 = |\vec x-\vec x_i|^2
\end{equation}
%
For $\Lambda=0$ these solutions reduce to the $D=5$ generalizations \cite{Myers:1986rx} of the Majumdar-Papapetrou multi-black hole solutions \cite{Majumdar:1947eu,Papaetrou:1947ib}, while if all the masses $M_i$ are set to zero, one obtains $D=5$ deSitter spacetime.  
For $N=1$, the case of a single center, the metric (\ref{basicmetric}) is diffeomorphic to the static Reissner-Nordstrom-DeSitter metric with charge $Q=\pm M$.  

Static, non-extreme, charged deSitter black holes have three horizons, inner and outer black hole horizons and a deSitter horizon.  For fixed charge $Q$ and cosmological constant $\Lambda$, the mass of the black hole is constrained to lie in a range  $M_{min}< M< M_{max}$.  In the extremal limits $M=M_{min}$ and $M=M_{max}$ the outer black hole horizon coincides respectively with the inner horizon or with the deSitter horizon.  For $M$ outside this range, the Reissner-Nordstrom-deSitter solution represents a naked singularity.

It is well known that for $\Lambda=0$, the solutions (\ref{basicmetric}) represent collections of extremal black holes with $M=M_{min}$. The degenerate black hole horizons are located at $\vec x =\vec x_i$ which can be seen from the metric to be spheres of non-zero area. For $\Lambda>0$ this is no longer the case.  The black holes are non-extremal, having distinct, non-degenerate inner and outer black hole horizons.  The spacetimes  in (\ref{basicmetric}) have the interesting property that the Hawking temperature of the black hole is equal to the deSitter temperature  \cite{Kastor:1992nn,London:1995ib}, both of which vanish in the limit $\Lambda=0$.  The black holes are thus in quantum mechanical equilibrium with their background. 

\begin{figure}[htbp]
\centering
\includegraphics[totalheight=2.5in]{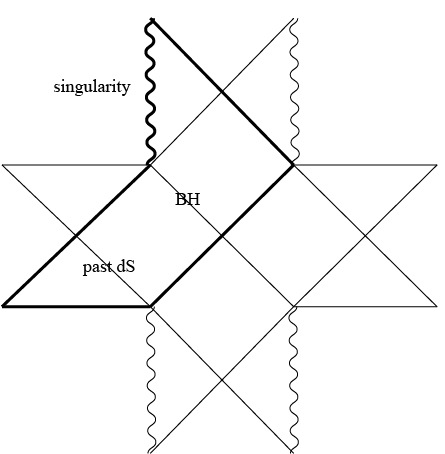}
\caption{{\it Conformal diagram for non-extreme Reissner-Nordstrom-deSitter black holes.  The bold lines outline the region covered by the deflating coordinate patch.  This includes the black hole and past deSitter horizons.  The wavy lines indicate the curvature singularity.}}
\label{deflatingpatch}
\end{figure}

In order to cover the entirety of deSitter spacetime, it requires both the inflationary coordinate patch with $H=+H_0$ and the deflating patch with 
$H=-H_0$.  Similarly, as described in \cite{Kastor:1992nn,Brill:1993tm}, both inflating and deflating patches are required to cover the entirety of the Reissner-Nordstrom deSitter spacetimes.  For $M_{min}<M<M_{max}$, the inflating patch covers the white hole horizon and the future deSitter horizon while the deflating patch covers the black hole horizon and the past deSitter horizon.  The region of the Reissner-Nordstrom-deSitter conformal diagram covered by the deflating coordinate patch is outlined by the bold lines in figure (\ref{deflatingpatch}).

For the multi-center solutions, the deflating patches similarly cover the black hole horizons.  However, now the black holes are embedded in a collapsing universe.  It is argued in \cite{Kastor:1992nn,Brill:1993tm} in the $D=4$ case that if one has a collection of $N$ black holes with masses $M_i$ such that $\sum_{i=1}^N M_i<M_{max}$ then the black holes merge, driven together by the collapse of the universe, into a single black hole with mass $M=\sum_{i=1}^N M_i$.  This conclusion is also supported by the numerical work of \cite{Nakao:1994mm}.

\begin{figure}[htbp]
\centering
\includegraphics[totalheight=1.2in]{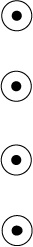}
\caption{{\it Spherical black hole apparent horizons surrounding each individual center in the array at early time with large scale factor $a(t)$.}}\label{arrayfigure}
\label{kkbhdsfig}
\end{figure}

We now present the metric for the system we are interested in: a single charged black hole in a collapsing Kaluza-Klein--deSitter background.
%
%
In order to obtain a black hole in our KKdS background with period $L$ we consider a one dimensional array of black holes also with period $L$ in the KT metric (\ref{basicmetric}). We are interested in what happens to the apparent horizons.  Before moving on to the explicit solutions, let us present our expectations.  If we take the deflating patch and look back far enough in time, then for each black hole in our array, we can find an apparent horizon of small radius surrounding it that corresponds to the black hole horizon for a single isolated $Q^2=M^2$ RNdS black hole as in figure  (\ref{arrayfigure}), as well as a spherical deSitter horizon surrounding each hole.  As the scale factor decreases, these spheres will increase in size and presumably distort.  We are interested in seeing what will happen as the scale factor decreases further.  Do the black holes merge into a black string, with the event horizon changing from spherical to cylindrical topology, or do they somehow remain separate, preserving the topology of the horizon? Does  the deSitter horizon evolve as in a KKdS spacetime
with no black hole (see section \ref{kkdssec} below), effortlessly changing its topology as the physical size of the compact dimension shrinks?

The metric for a one dimensional array of black holes described by (\ref{basicmetric})  with mass $M$ and separated by $L$ along the $w$ axis is given by
\begin{eqnarray}\label{metric}\label{kkmetric}
ds^2&=&-{ dt^2 \over U^2}+ a^2\,U\left[dw^2+d\rho^2+\rho^2\,d\Omega_{(2)}^2 \right]\\ 
\label{lattice}
U&=& 1 +{f(w, \rho ) \over a^2}, \qquad f= M \sum_{n=-\infty}^\infty{1\over  \rho^2 +(w+nL)^2 }.
\end{eqnarray}
We will consider only the deflating coordinate patch 
with $a(t) = \exp{(-H_0t)}$ and $H_0 >0$, which in the case of an isolated RNdS black hole covers the black hole and past deSitter horizons.

Such a periodic array was consider in the  asymptotically flat case in reference \cite{Myers:1986rx}  and the sum here can be carried out in the same way\footnote{See also \cite{Harmark:2002tr} for a detailed treatment of such sums.} giving the result
%
\begin{equation}\label{fdef}
f(w,\,\rho)=\frac{\pi\,M}{L\,\rho}\,\frac{\sinh(2\pi\,\rho/L)}{\cosh(2\pi\,\rho/L)-\cos(2\pi\,w/L)}
\end{equation}
which clearly displays the periodicity $w\equiv w+L$ in the compact direction.  

The KKdS black hole spacetime  (\ref{kkmetric}) has an overall $SO(3)$ rotational symmetry in the non-compact spatial directions.
In the far field limit where the transverse radial coordinate $\rho>>L$, one finds that the metric function becomes  
$f\simeq \pi M/\rho L$, displaying the overall 
$1/\rho$ fall-off characteristic of a finite mass object in $4$ noncompact dimensions. In the near field limit with $\rho,w\ll L$ we retain the two leading terms in the expansion 
\begin{equation}\label{smallcoord}
f\simeq {M\over \rho^2+ w^2} + {\pi^2M\over 3\,L^2} +{\cal O}(w^2,\rho^2)
\end{equation}
which display an overall $SO(4)$ invariance that is broken by the subsequent term.  The second term in (\ref{smallcoord}) gives a contribution to the metric function at the origin coming from the entire one dimensional array of charges.

\section{Apparent horizons}\label{apparent}

We are interested in studying event horizons in the KKdS black hole spacetimes.  However, these spacetimes are not static, and the locations of the event horizons, which depend on the entire evolutionary history of the spacetime, are not easily found.  
It is more feasible to look for apparent horizons, which are defined in terms of the data only on a spatial slice, and trace out there evolutions.
An apparent horizon is a closed surface on a spatial slice on which the expansion $\theta$ of null rays normal to the surface vanishes.  If $\Sigma$ is a spatial slice with induced metric $h_{ab}$ and extrinsic curvature $K_{ab}$, and $S$ is a closed surface on $\Sigma$ with unit normal $n^a$ then the expansion of null rays normal to $S$ is given by
\begin{equation}\label{expansions}
\theta =D_a\,n^a-K_{ab}\,n^a\,n^b+K,
\end{equation}
where $D_a$ is the covariant derivative operator on $\Sigma$ compatible with the induced metric $h_{ab}$ and $K=h^{ab}K_{ab}$.  Note that the normal $n^a$ can be chosen to point either inward or outward from $S$.  We could then define expansions $\theta _\pm $ corresponding to  either ingoing or outgoing null rays respectively.  However, we will be mainly interested in the expansion of outgoing null geodesics, and so will drop the subscript and write $\theta \equiv \theta_+ $.  Apparent horizons for KT-spacetimes with a pair of black holes were studied in reference \cite{Nakao:1994mm}\footnote{See also reference \cite{Nakao:1992zc} for a related study of apparent horizons of initial data with multiple Einstein-Rosen bridges and positive cosmological constant}.


Since we will be looking for horizons in the rather complicated Kaluza-Klein-deSitter black hole 
spacetime, let us first fix our intuition by studying two limiting cases: a single Reissner-Nordstrom-deSitter black hole with no compact dimension, and Kaluza-Klein-deSitter spacetime with
no black hole.

\subsection{Isolated RNdS black hole}\label{isolated}

We start by considering apparent horizons of  a single, isolated $D=5$  RNdS black hole with $Q^2=M^2$.   These objects are the building blocks for constructing the localized KKdS black holes and the horizons of the KKdS black holes will be well approximated by the single black hole results at early times when the size of the compact direction is large.  The single black hole spacetimes are static and therefore the apparent horizons coincide with the event horizons.  In order to compute the expansion for radially outgoing null rays using equation (\ref{expansions}), we need to assemble a number of ingredients. The metric function for a single black hole of mass $M$ at the origin is given by $U = 1 + M/a^2 r^2$, the outgoing unit normal to a sphere has  $n^r =1/(a\,U^{1/2})$ and the extrinsic curvature of a surface of constant cosmological time $t$ is $K_{ab}= H h_{ab}$.  Combining these ingredients results in the expression for the expansion
\begin{equation}\label{expansions1}
\theta =3\left( {1\over a\,r\,U^{3\over 2}}+  H\right)
\end{equation}
Since the metric function $U(r)$ is positive, one sees immediately that solutions
$\theta=0$ exist only if $H =-H_0 $ is negative, {\it i.e.} in a contracting deSitter phase.  This  backs up the statement made earlier  that is the deflating coordinate patch that covers the  black hole horizon\footnote{For ingoing null rays, the sign of the first term in (\ref{expansions}) is reversed and one finds that the white hole horizon is covered by the inflating coordinate patch.}.

It is useful to rewrite the equation $\theta=0$ for the black hole and past deSitter horizons in the form
\begin{equation}\label{apphor}
{y\over (1+y)^{3/2}}=\epsilon^{1/2}
\end{equation}
with $y=a^2\,r^2/M$ and $\epsilon=MH_0^2$.  A straightforward analysis of the function on the left hand side shows that solutions only exist when the parameter $\epsilon\le 2/3^{3\over 2}\simeq 0.4$, with equality corresponding to the extremal limit in which the black hole and deSitter horizons coincide.  The solutions of (\ref{apphor}) can be found approximately in the limit that $\epsilon\ll 1$, with one solution occurring each in the regimes of small and large values of the dimensionless coordinate $y$.

The small $y$ solutions to (\ref{apphor})  are given at leading order in $\epsilon$ simply by $y=\epsilon^{1/2}$.  Translating back into the comoving radial coordinate and including the next order term, one has
\begin{equation}\label{bhhorizon1}
r_{BH}=\frac{M^{1/2} }{a}\,\epsilon^{1/2}\left(1+{3\over 4}\epsilon^{1/2}+{\cal O}(\epsilon)\right).
\end{equation}
One can confirm that this is a black hole horizon by checking that the expansion goes from negative values on the inside to positive values outside.
For large $y$ the solution is given at leading order by $y=\epsilon^{-1}$, which along with the next correction translates into
\begin{equation}\label{dshorizon}
r_{dS}=\frac{M^{1/2} }{a}\,{1\over \epsilon^{1/2}}\left(1-{3\over 2}\epsilon+{\cal O}(\epsilon^2)\right).
\end{equation}
In this case, one can check that the expansion changes from positive to negative in moving outward through the horizon, as it should for a past deSitter horizon.  One sees that to leading order in this regime $r_{BH}=\epsilon\ r_{dS}$, so that the horizons are indeed well separated.  

One also sees that the time dependence of both horizons is simple.  The products of the scale factor and the comoving horizon radii are constants.  The universe is collapsing.  The scale factor $a(t)$ is large at early times and shrinks to zero at late times.  Both the black hole and deSitter horizons vary inversely with the scale factor and are therefore small spheres at early times, expanding out to infinity as the collapse proceeds.

 
The  areas of the black hole and deSitter apparent horizons are found to be constant in time, corresponding to the fact that for a single RNdS black hole they coincide with the event horizons.  They are given approximately by
 \begin{equation}\label{notkkarea}
A_{BH} \simeq \Omega_3 M^{3/2}\left(1+{3\over 2}\epsilon^{1/2}\right), \qquad A_{dS}\simeq {\Omega_3 \over H_0 ^3}(1-3\epsilon)
\end{equation}
where $\Omega_3=2\,\pi^2$ is the area of a unit $3$-sphere.
%
The signs of the correction terms illustrate the general feature that the black hole horizon is expanded from its asymptotically flat size by the positive cosmological constant, while the deSitter horizon is pulled in by the presence of the black hole.
 

It is also of interest to understand the character of the hypersurfaces foliated by apparent horizons on successive time slices, whether they are null, timelike or spacelike.  In the present case, we know that the apparent horizons coincide with the cross-sections of event horizons and therefore sweep out null hypersurfaces.  However, this is not at all obvious from the cosmological form of the metric.  Moreover with applications to non-static spacetimes in mind, it will be useful to see how we can ascertain the character of a family of apparent horizons whose positions are time dependent.
%

Denoting time derivatives by a dot, it follows from $\dot{a} = -a\,H_0$ that the rate of change of the deSitter horizon radius is given by
\begin{equation}\label{dsdot}
\dot{r}_{dS} = H_0\ r_{dS}
\end{equation}
Note that this is positive because $\dot{a}$ is negative.
The sign will reverse for the inflating cosmology.  Radially directed null
rays propagate according to 
\begin{equation}\label{nulldot}
\dot{r}_\pm = \pm {1\over a\,U^{3/2}}
\end{equation}
where the $+$  sign is for outgoing rays, and the $-$ is for ingoing. When the null ray is at an apparent horizon equation (\ref{expansions}) holds, so the right hand side of (\ref{nulldot}) can be rewritten as 
\begin{equation}\label{nulldothor}
\dot{r}_\pm = \pm H_0\ r
\end{equation}
One sees that  ingoing null rays can cross the expanding deSitter horizon (\ref{dsdot}). However, outgoing null rays just stay with the deSitter horizon, but cannot cross is to larger $r$.  The surface swept out by the deSitter apparent horizons is then seen to be null.  A similar result obviously holds for the black hole apparent horizon.


%

\subsection{Kaluza-Klein-deSitter spacetimes}\label{kkdssec}

Another interesting limit to consider is the KKdS spacetime itself, with no black holes.
We start by defining $D=5$ KKdS spacetime as a compactification of $D=5$ deSitter spacetime along a spatial symmetry direction by writing deSitter metric in flat cosmological coordinates
\begin{equation}\label{kkds}
ds^2 = -dt^2 + e^{2Ht}\left(dw^2 + d\rho^2 + \rho^2d\Omega_{(2)}^2 \right)
\end{equation}
and making the identification $w\equiv w+L$.  The uncompactified deSitter spacetime has a static
Killing vector, timelike inside the deSitter horizon, that is given in cosmological coordinates by
\begin{equation}\label{killingvector}
\xi^a = (\partial /\partial t){}^a - Hw(\partial /\partial w){}^a - H\rho(\partial /\partial \rho){}^a.
\end{equation}
%
This expression is not invariant under the identification $w\equiv w+L$ and therefore the KKdS spacetime fails to be static.  The periodic boundary conditions in the $w$ direction break the local symmetry generated by the vector field (\ref{killingvector}). 

\begin{figure}[htbp]
\centering
\includegraphics[totalheight=1.5in]{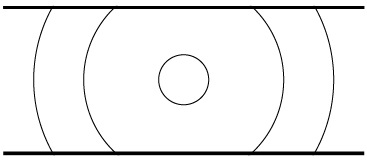}
\caption{{\it Expanding $S^3$ apparent horizons in KKdS.  The horizontal lines at top and bottom are at $w=\pm L/2$ respectively and are identified.}}
\label{kkdsfig}
\end{figure}

We now turn to finding apparent horizons. Since the spacetime is still locally deSitter, this starts out as a simple task.
The expanding $3$-spheres found in equation (\ref{dshorizon}) in the  limit  that the parameter $\epsilon$ is taken to zero give 
the usual expression for the deSitter horizon radius in comoving coordinates  $r_{dS}=1/aH_0$.  However, we need to consider how these apparent horizons are affected by the periodic identification of the KKdS spacetime.  If we represent the KK deSitter spacetime as in figure (\ref{kkdsfig}), then at early times there are spherical apparent horizons having comoving radius less than $L$.  As the scale factor shrinks these spheres expand and run into the identifications $w\equiv w+L$.  The 
$\theta =0$ surface then makes a topological transition to a non-smooth, or ``scalloped",  cylinder that gradually flattens out as it expands. We will refer to these as spherical arcs.
Together, the  spheres and spherical arcs sweep out a  null  surface which is a past deSitter horizon, as in the uncompactified spacetime. Signals can enter from larger $r$ to the inside, but can not travel from the inside to the outside.  In the KKdS case, however, the deSitter horizon makes a transition from spherical to (non-smooth) cylindrical topology.
The spheres have constant area $A_{dS, sphere} = 2\pi^2\,H_0 ^{-3}$ until they become cut-off into arcs. The area is then given approximately by
$A_{dS, arc}=4\pi\, L \,a/H_0^2$ in the regime with $a\la 1/L\,H_0$, and so the area decreases along with the scale factor.

The story is complicated by the fact that KKdS spacetimes also have smooth apparent horizons with cylindrical topology and exact cylindrical symmetry.  If we consider cylinders centered on the line $\rho=0$ and translationally invariant in the $w$-direction, then the unit outward normal has
 $n^\rho=1/a$.  One finds that the $\theta=0$ surfaces in the deflating universe are given by
\begin{equation}\label{cylindrical}
\rho_{dS} = {2\over 3\,a\,H_0}.
\end{equation}
We see that these cylindrical apparent horizons expand more slowly in radius than the spherical apparent horizons described above,
and so sweep out a timelike, rather than null, hypersurface. Hence these cylinders do not play a role in delineating the causal structure of the KKdS spacetime and we see that even in this very simple example there are natural, symmetrical, apparent horizons that are not relevant for mapping out the causal structure of the spacetime.   We will need to keep this in mind in our discussion of the apparent horizons of the localized KKdS black hole.

We note that, if one does a Kaluza-Klein reduction of (\ref{kkds}) to four dimensions, one can check that the time dependence (\ref{cylindrical}) is also characteristic of the apparent horizons in the four-dimensional Einstein metric that results.  This metric is not static, so it is consistent that its apparent horizon has a time dependent area.  It is also possible to find a discrete family of smooth apparent horizons of KKdS spacetimes having cylindrical topology, but not cylindrical symmetry.  These are higher dimensional versions of constant mean curvature  Delaunay cylinders \cite{delaunay}.

\section{Apparent horizons for localized KKdS black holes}\label{kkdshorizons}

We are now prepared to look for  apparent horizons on constant time surfaces of the localized KKdS black hole spacetime given in (\ref{metric}). The extrinsic curvature of a constant time slice is $K_{ab}=-H_0 \,h_{ab}$ where the induced spatial metric $h_{ab}$ is simply the spatial part of the full metric.
Given that $h_{ab}$  is spherically symmetric in the non-compact spatial dimensions, 
we will look for apparent horizons that share this symmetry.  Such surfaces may be specified by a function $\rho(w)$ giving the dependence of the non-compact radial coordinate of the surface on the coordinate along the compact direction. The in- and outgoing unit normals are then given by
\begin{equation}
n_{\pm a}dx^a =\pm \left( {a^2 U\over 1 +\rho^{\prime 2}}\right)^{1\over 2}(-\rho^\prime \,dw +d\rho)
\end{equation}
where $\rho^\prime = d\rho/dw$ and the corresponding expansions $\theta_\pm$ are found from (\ref{expansions}) to be
\begin{equation}\label{theta}
\theta_\pm=\pm\frac{1}{\sqrt{1+\rho'{}^2}}\,{1\over a\,U^{1/2} }\,
\left[\frac{3}{2\,a^2\,U}\,( \partial_\rho f -\rho'\,\partial_wf )
+\frac{2}{\rho}-\frac{\rho''}{1+\rho'{}^2}\right] -3\,H_0 .
\end{equation}
As discussed above we will restrict our attention to the parameter regime $\epsilon\ll 1$ 
where the mass scale is small compared to  the deSitter scale. 
We expect that at early times when the proper length of the $S^1$ is large, there will be a  nearly spherical black hole horizon fitting well inside a nearly spherical cosmological horizon.

We first summarize our findings\footnote{
These results apply as the time coordinate flows from  $- \infty < t < \infty$ and correspondingly  the scale factor flows in the range  $ \infty > a^2(t) > 0$.
The metric can also be continued to negative values of $a^2$ by the simple substitution 
$a^2 =-H_0 \tau$, $- \infty < \tau < \infty$.  This is necessary to study the portion of the spacetime
that contains the singularities \cite{Brill:1993tm}. However, in this paper we will restrict ourselves
to positive values of $a^2$.}.
\begin{enumerate}

\item  {\bf Black hole apparent horizons:}  We find that the curvature singularity at $r=w=0$ is always surrounded by a very nearly spherical black hole horizon whose area stays approximately constant in time with $A_{bh}\simeq M^{3/2}\,\Omega_{(3)}$.
Contrary to intuition, the size of the compact dimension does not get small in the neighborhood of the black hole and the horizon does not change topology.

\item {\bf DeSitter apparent horizons:}  In this case we find several solutions that evolve as follows.

\begin{enumerate}

\item {\bf Early time:} In the regime $a(t)\ga 1/ LH_0$ the black hole horizon is surrounded by a nearly spherical dS horizon, whose area is also approximately constant in time with $A_{sph, dS} =H_0^{-3}\,\Omega_{(3)}$. This sphere is a past dS horizon, {\it i.e.} light signals enter from the exterior to the interior, but cannot cross from inside to outside.

\item {\bf Intermediate times:} The deSitter spheres have comoving radius larger than $L$, and so become cut-off spheres which are topological cylinders. We will refer to these surfaces as spherical arcs. They  sweep out a  past dS horizon with an area that decreases in time in proportion to the decreasing size of the compact dimension, $A_{arc, dS} =a\,L\,\Omega_{(2)} / H_0^{2}\ $. Note that the cases for early and intermediate times show the same behavior as the deSitter horizons as in the KKdS spacetime with no black hole.

\item {\bf Late times:} In the regime $a(t) \la M H_0 /L $ the boundary defined by the evolution of the deSitter arcs becomes a future deSitter horizon.  
The area  of the horizon asymptotes to a constant value that depends on both $M $ and $H_0$, $A_{arc, dS} =\Omega_{(2)} M / H_0$. This horizon is distinct from any of the horizons in either the KKdS spacetime or extended RNdS.

\item {\bf Cylinders:}  As was the case for the pure KKdS spacetime described above, there are also smooth
cylindrical deSitter apparent horizons that do not have appear to be associated with causal boundaries.

\end{enumerate}

\end{enumerate}
 Remarkably, we find no evidence of black hole horizons with cylindrical topology. We will discuss the origin and the physical significance of this fact in the next section.

\subsection{Spherical horizons}\label{sphhor}
We first look for small, approximately spherical horizons which we expect to exist in the early time regime of large scale factor. If we assume that the quantities $2\pi\,w $ and  $2\pi\,\rho$ are both much smaller than $L$, then the function $f$ can be approximated as 
\begin{equation}\label{smallr}
f \simeq\,\frac{M}{w^2+\rho^2}+\frac{\pi^2\,M}{3\,L^2 }+{\cal {O}}\left(\rho^2/L^2,\,w^2/L^2\right).
\end{equation}
Here we keep the next to leading term explicitly because it preserves the full $SO(4)$ rotational symmetry of the leading term.   Subsequent terms in the expansion break this symmetry.  While the leading term in equation (\ref{smallr}) reproduces the behavior of the isolated RNdS black hole, the constant term arises from the one dimensional array of KK images of our black hole.   As we will see this term plays an important role in determining the late time behavior of the small, approximately spherical black hole horizons.

Within the approximation (\ref{smallr}) we can now look for solutions that have exact spherical symmetry, with $\rho ^2 (w) = R_h ^2 - w^2 $ and $R_h$ constant.   To the extent that these solutions have $2\pi R_h\ll L$, the solutions to the full equations will be very nearly spherical in shape.
The apparent horizon equation $\theta=0$ can now be written as
\begin{equation}\label{sphtheta}
a\,H_0\,R_h\,U^{1/2}  = 1 +{1\over 2\,a^2\,U} (\rho\,\partial_\rho f + w\,\partial _w f )
\end{equation}
with the relation $\rho ^2+w^2 = R_h ^2$ assumed.
Substituting  in the approximate expression (\ref{smallr}) for $f$ this partial differential equation
for $\rho (w)$ becomes an algebraic equation for $R_h$. If one now defines the variable
\begin{equation}\label{newy}
y= {R_h ^2 \over M}( a^2 + \frac{\pi^2 M}{3L^2 })
\end{equation}
then the resulting algebraic equation for $y$ is identical to equation (\ref{apphor}) above, which arose in the context of a single, isolated RNdS black hole.
The analysis of solutions to this equation in terms of $y$ is unchanged.  For $\epsilon\ll 1$ there is one solution at small $y$ and one at large $y$ given by $y\simeq \epsilon^{1/2}$ and $y\simeq 1/\epsilon$ respectively.  

What has changed, however, is the relation between the variable $y$ and the apparent horizon radius $R_H$.  The case of a single, isolated RNdS black hole, as discussed in section (\ref{isolated}), can be recovered from (\ref{newy}) by taking the $L\rightarrow\infty$ limit which eliminates the second term in the parenthesis.  In any case at early times the $a(t)^2$ term in (\ref{newy}) dominates and the result is the same as for RNdS.
The black hole and deSitter horizons will be small spheres that expand outward as the scale factor decreases, as in equations (\ref{bhhorizon1}) and (\ref{dshorizon}).  
In the late time limit, on the other hand, the second term in parenthesis in (\ref{newy}) dominates and the behavior is entirely changed from the isolated black hole case.  In this limit both the black hole and deSitter horizon radii asymptote to constant values.  

We note that the area of both types of horizons stay constant within our near field approximation.  Any sphere  of radius $R$ with $R\ll L$ has area given by
\begin{equation}\label{areasph}
 A_{sph} = a^3\,U^{3/2}\,R^3\,\Omega _{(3)}\,,
 \end{equation}
If the sphere is an  apparent horizon, then using equation (\ref{apphor}) for an apparent horizon, the area can be written in the simple form
\begin{equation}\label{areasphtwo}
A_{sph,hor} \simeq  \frac{M}{H_0}\,y\,\Omega _{(3)}
\end{equation}
where $y$ is a root of (\ref{apphor}), which is manifestly constant in time. 

\underline{\bf{Black Hole Horizons}}

We expect that the nearly spherical apparent horizons with $y\simeq\epsilon^{1/2}$ are of black type as in the isolated black hole in section (\ref{isolated}).
Plugging in for the variable $y$ from (\ref{newy}) one finds that the horizon radius is given by
\begin{equation}\label{bhradius}
R_{BH}^2 \simeq {M\epsilon^{1/2}\over a^2 +{\pi^2 M\over 3 L^2}},\qquad
\end{equation}
%
One can check that this surface is a black hole type horizon by restoring the inequality sign in the trapped surface analysis. The condition $ \theta \geq 0$ becomes $ \epsilon^{1/2}\,(1+ y)^{3/2} \leq y $.
In the regime  $y\ll 1$, we see that $\theta$ is positive for $y \geq \epsilon^{1/2}$, confirming the identification.  

As noted above, for early times the evolution of the black hole horizon matches on to that of the isolated RNdS black hole, starting from small size and expanding as the universe collapses.  However, rather than expanding forever, at late times as the scale factor $a(t)$ goes to zero the square of the horizon radius asymptotes to the finite value $R_{BH}^2\simeq 3\,L^2\,\epsilon^{1/2}/\pi^2$.   One can check that if the parameter $\epsilon$ is taken to be sufficiently small, then throughout its entire evolution the horizon satisfies $R_{BH}/L\ll 1$.   It then follows that the approximate form of the function $f(\rho,w)$ in (\ref{smallr}) is valid throughout and therefore that (\ref{bhradius}) represents an approximate solution over $-\infty<t<\infty$.  We see that this family of apparent horizons never self-merges around the compact dimension, but rather stays spherical  to good approximation and expands only out to a finite size that is smaller than $L$.


Next we show that the family of apparent horizons $R_{bh} (t)$ sweeps out a 
null surface that is tangent to outgoing radial null geodesics. 
To see this, note that
ingoing and outgoing transverse radial null geodesics propagate according to
\begin{equation}\label{nullgeoeq}
\dot{\rho}_\pm   =  \pm {1\over a\,U^{3/2}} 
\end{equation}
On the path of the photon the coordinates $\rho(t)$ and $R(t)$ coincide.  Evaluating the right hand side  of (\ref{nullgeoeq}) on $R_{bh}$ gives
\begin{equation}\label{rbhdot}
\dot{\rho}_{\pm} =\pm a^2 {M ^{3/4} H_0 ^{3/2} \over (a^2 +{\pi ^2 M \over 3L^2} )^{3/2} }
\end{equation}
On the other hand, taking the time derivative of $R_{bh}$ gives equation (\ref{rbhdot}) with the plus sign;
the apparent horizon propagates along an outward going null ray.
 Ingoing null rays can cross $R_{bh}$ from large $\rho$ to smaller $\rho$ but
 outgoing rays cannot exit; it is a black hole horizon.
 
 Interestingly, within the approximation given by (\ref{smallr}), since the black hole apparent horizon is always in the near field regime, it has constant area given by 
\begin{equation}\label{areasphbh}
A_{sph, bh} =\Omega _{(3)} M^{3/2} 
\end{equation}
This is not  surprising at early times, when the system is effectively decompactified.   However, the spherical black hole horizon with this area survives at late times because in the regime where (\ref{smallr}) holds the KKdS black hole metric is equivalent to a {\em single} RN black hole on a {\em non expanding background}. To see this, first observe at late times with $f/a^2\gg 1$
 the KKdS black hole metric (\ref{kkmetric}) is approximately
\begin{equation}
ds^2\simeq -\frac{dt'{}^2}{f(w,\rho)^2}+f(w,\rho)\,\left(d\rho^2+dw^2+\rho^2\,d\Omega_{(2)}^2\right)\,\,,
\end{equation}
where $dt'=a^2 dt$.
This metric is time-independent. In the near field
regime where $f$ is approximately given by (\ref{smallr}) 
we can further redefine $\rho$ and $t'$ to bring the metric to the  form
\begin{equation}\label{nearbh}
ds^2=-\frac{d\bar{T}^2}{\left( 1+M/\bar{R}^2\right)^2}+\left(1+\frac{M}{\bar{R}^2}\right)\,\left(d\bar{R}^2+\bar{R}^2\,d\Omega_{(3)}^2\right)\,\,,
\end{equation}
valid for  ${\bar{R}}\ll \sqrt{M}$. This is  the metric of a single $D=5$ RN black hole in isotropic spatial coordinates, a static spacetime
with a spherical horizon of constant area.  One can check that this behavior is not specific to an infinite array of back holes.   Two black holes are already sufficient to give equation (\ref{nearbh}). The second black hole provides a potential at the location of the first one, regularizing the behavior of the metric in the limit that the scale factor $a(t)$ shrinks to zero. 
The approximate metric near the black hole horizon is static with Killing vector  ${\partial \over \partial \bar{T}}$ becoming null at $\bar{R} =0$. The area of  this horizon is $A_{bh} = \Omega _{(3)} M^{3/2}$, in agreement with (\ref{areasphbh}).

Note that as a consequence of the above argument, the size of the KK circle  in a neighborhood of the KKdS black hole  {\em does not} shrink. Even if the compact direction {\em does} shrinks in the far field, the KKdS black hole horizon does not  evolve into a cylinder.  One might expect that a new, cylindrical black hole horizon could form outside the spherical black hole horizon described above.  Such an expectation would be motivated by the analytical work of~\cite{Brill:1993tm} as well as the numerical studies of~\cite{Nakao:1994mm}. In those works, it was shown that two RN black holes in a collapsing deSitter spacetime maintain their horizon and, when they get closer than a certain critical distance, develop a second apparent horizon that encircles both of them, signaling their coalescence. However,  in our case both numerical and analytical  work shows that such a second BH horizon with cylindrical topology does {\em not} form. We conjecture that this is due to the fact that the end state of the system would correspond to a singular string, and that our regular system does not evolve into such a singular final state. This will be examined in more detail in Section~\ref{string}.

\begin{figure}[htbp]
\centering
\includegraphics[totalheight=2.5in]{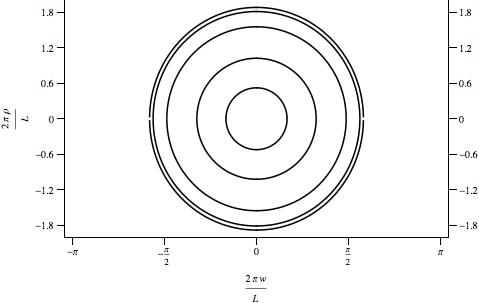}
\caption{{\it Evolution of the comoving radius of the spherical black hole horizons for the choice of parameters $H=0.2\times(2\,\pi/L)$, $M=(L/2\,\pi)^2$. From the innermost to the outermost horizon, the scale factor $a$ takes the values $1,\,5^{-1/2},\,5^{-1},\,5^{-3/2}$~and $5^{-2}$. Note that according to eq.~(\ref{bhradius}), the radius of the black hole horizon should converge as $a\to 0$ to $\sim 1.88\times (L/2\,\pi)$.}}
\label{bhhor}
\end{figure}

\underline{\bf{DeSitter Horizons}}

We now turn to the nearly spherical apparent horizon with $y\simeq 1/\epsilon$, which  will correspond to a deSitter type horizon.  From (\ref{newy}), we have that
\begin{equation}\label{cossphere}
R_{dS}^2 \simeq {M\over \epsilon\,(a^2 +{\pi^2 M\over 3\,L^2})}\, .
\end{equation}
Like the black hole apparent horizon (\ref{bhradius}), this expression for the deSitter horizon radius starts out small at early times and asymptotes to a constant value at late times.  However, in this case the late time constant value has $R_{dS}^2/L^2\simeq 3/(\pi^2\epsilon)$ which is much greater than one in the limit of small $\epsilon$.   Hence, unlike in the black hole case, the condition $R\ll L$
does not hold over the entire range of $a(t)$ in (\ref{cossphere}).  Therefore,  (\ref{cossphere}) is a good approximate solution only for early times with $a\gg 1/H_0 L$.  Our analytic approximation breaks down at this point and the fate of the deSitter apparent horizon must be studied by other means.

%

One sees that $R_{dS}$ is indeed a  cosmological or deSitter  horizon by restoring the inequality in the trapped surface equation, which shows that $\theta$ is positive for $R< R_{dS}$ and negative for $R>R_{dS}$. As a check,  note that at early times when $a$ is very large the horizon radius approaches the value   $R_{dS}\simeq \left[1-\pi^2\,M/(6\,L^2\,a^2)\right] /(a H_0 ) $.  The leading term agreeing with the deSitter horizon in RNdS. The next correction makes the radius  smaller than $1/(a  H_0 )$. This is similar to the case of the single RNdS black hole~(\ref{dshorizon}), where the effect of the mass is to pull the horizon in. However for the RNdS black hole the correction is proportional to $\epsilon=MH_0^2$, whereas in the present case it is proportional to $M/(a^2\,L^2)$, indicating that the dominant effect comes from the compact nature of the space.
         
The early time deSitter spheres have  the same causality behavior as the KKdS spacetime case with $M =0$ that was analyzed in Section (\ref{kkdssec}). The calculations are almost identical, since $U \approx 1$ in the regime $y\gg 1$. Hence $R_{dS}$ is a {\em past} deSitter horizon -- light signals can enter from larger $\rho$ to the interior, but cannot go from the inside to the outside. The area of the early time spherical cosmological horizon is given by
\begin{equation}\label{areasphcos}
A_{sph, dS} =\Omega _{(3)} {1\over H_0 ^3 }  
\end{equation}
The area is constant as long as the sphere has comoving radius small compared to $L$,
as demonstrated in the beginning of this section.

\begin{figure}[htbp]
\centering
\includegraphics[totalheight=2.5in]{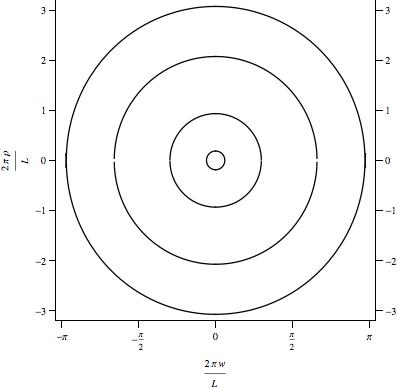}
\caption{\it Evolution of the spherical cosmological horizons for the choice of parameters $H=0.2\times(2\,\pi/L)$, $M=(L/2\,\pi)^2$. From the innermost to the outermost horizon, the scale factor $a$ takes the values $25,\,5,\,5^{1/2},\,5^{1/4}$.  The horizontal axis extends only up to $2\,\pi\,w/L=\pi$, since $-L/2<w<L/2$. For $a\lesssim 5^{1/4}$ the radius of the horizon becomes larger than the compactification radius. Note also that, even for the outermost horizon when the approximation $2\pi\,R_h\ll L$ is not valid any more, the numerical solution still follows closely the analytical behavior described in the main text. }
\label{sphecos}
\end{figure}

\subsection{Boundaries of Self Intersecting Spheres: deSitter Arcs}

We have seen that the approximate solution (\ref{bhradius}) 
holds for the entire time evolution of the spacetime and describes a black hole of constant area.
However, equation (\ref{cossphere}) describing the spherical deSitter apparent horizons is only valid at early times with $a\, R_{dS}\lesssim L$.
Therefore the evolution of the deSitter apparent horizon beyond the early time is yet to be determined.  

We have studied the exact apparent horizon equations in detail  using the full expression (\ref{fdef}) for the function $f(\rho,w)$ via numerical methods.   In the case of the black hole horizon, this study confirms the conclusions of our analytic work and also shows that no cylindrical black hole type apparent horizon forms surrounding the limiting sphere.  

However, in the deSitter case our numerical study shows that the expansion of the deSitter apparent horizon continues throughout the spacetime evolution.  At early times the radius matches on to (\ref{cossphere}) and then it continues to grow with time.  As the deSitter spheres grow beyond the regime in which (\ref{smallr}) holds, they lose exact spherical symmetry, but the spherical topology continues until they hit the compactification length $L$.  At this point, the evolution of the deSitter horizon continues in the form of spherical arcs, or scalloped cylinders, as discussed in section (\ref{kkdssec}).  These arcs flatten out as they continue to expand and eventually enter the far field regime with $\rho\gg L$ where they again match on to analytic results.  Both the analytic and numerical results demonstrate that the deSitter arcs asymptote to a fixed radius as we will next show.


The analytic results in the far field come about in the following way.  We look for a spherical solution in the {\em far} field where $\rho\gg L$. Hence we use the apparent horizon condition (\ref{sphtheta}) for spheres, but use the far field expression for $f$, $f\simeq \pi\,M /(L\,\rho)$, and the approximation that the constant radius $R_h$ is given by 
\begin{equation}\label{spharc}
R_h ^2 =w^2 + \rho^2 \approx \rho^2\, .
\end{equation}
This last approximation works because the  boundary that we are looking for only includes the part of the sphere where $\rho \gg w$.  With these approximations and setting  $ z=a^2\,L\,R/(\pi\,M )$, the equation $\theta =0$ becomes
%
\begin{equation}\label{arcthetwo}
\frac{1+2\,z}{ z^{1/2} (1+z )^{3/2}}={2\pi\,\epsilon \over H_0\,a\,L}
\end{equation}
which can be solved approximately in the two limiting regimes $z\ll 1$ and $z\gg 1$.  An approximate solution with $z\gg 1$ requires that the 
scale factor $a(t)$ lie in the intermediate regime with $\pi\,\epsilon\ll H_0\,a\,L\ll1$ and is given by $R_{dS,inter}  \simeq 1/ a H_0$.  
Hence in this intermediate time regime, the deSitter horizon continues to expand as if there is no compact direction, even though $R_{dS}>L$.
An approximate solution with $ z\ll 1$ holds at sufficiently late times such that  $H_0aL\ll2\pi\epsilon$ and behaves as
%
\begin{equation}\label{dsarctwo}
R_{dS,late} 
 \simeq {L\over 4\pi\epsilon }
\left[ 1+  \left( {H_0aL \over  2\pi\epsilon }\right)^2 \right] 
\end{equation}
%
%
We see that the deSitter horizon ultimately does asymptote to a constant radius in the late time, far field regime.  Note also that the time evolution of the horizon has also turned around between the intermediate and late time regimes, with $R_{dS,late}$ decreasing to its constant value.  We will see below that the causal nature of this late time limit of the deSitter apparent horizon is also surprising.

Recall that in the early time limit, the area of the deSitter apparent horizon is approximately constant with the value (\ref{areasphcos}).  The area of the far field spherical arcs, however, is not constant in time, shrinking due to the contraction of the compact direction until it ultimately reaches a finite value as the scale factor shrinks to zero.
For the surfaces (\ref{spharc}) we have 
\begin{equation}\label{areaarcs}
A_{arc}=4\pi\,a^3\,L\,R_h ^2\,U^{3/2}\end{equation}
Here a factor of $a(t)\,L$ comes from the integration around the compact direction. 
Respectively in the intermediate and late time regimes discussed above one finds
\begin{eqnarray}\label{medareaarc} 
A_{dS,inter} &=& 4\pi ^2\,\frac{a\,L}{H_0^2}\\
\label{lateareaarc} 
A_{dS,late} &=& 2\pi ^2 \,{M \over H_0 } \left[ 1 + {1\over 2}\left( {H_0aL \over  2\pi\epsilon }\right)^2\right].
\end{eqnarray}


The  finite limiting area for the deSitter apparent horizon is a surprising feature of our results. Since the limiting area is proportional to $M/H_0$ one sees that this behavior is a result of
the interaction of the cosmological constant and the black hole. Even more intriguing is
 the fact that area of the deSitter horizon $decreases$ to this value. In this spacetime, the
 area of the deSitter horizon does not have a simple interpretation as an entropy - it decreases with
time and so appears to violate the second law. Further, since the area of the black hole horizon
is constant, the total horizon area also decreases with time. 
One might wonder if there is another contribution to the total entropy of the spacetime leading to a generalized second law, but we leave this question for further work. On the other hand, the decrease in the area is a direct result of the decrease in the scale factor. One could say that the change in the area goes the same way as  the ``cosmological arrow" but goes in the opposite direction to the ``arrow of time," where the direction of time is set by the direction of propagation of geodesics of test particles.

A second related development, also surprising, is that the causal significance of the deSitter apparent horizon changes in the transition between the intermediate and late time regimes.
Specifically, while during the early and intermediate regimes the deSitter apparent horizon is  a past deSitter horizon, we find that by the late time regime it has evolved into a future deSitter horizon.  Null rays can leave, but not enter.
One can see this in the following way.  Note that the time derivative of $R_{dS}$ in the late time regime (\ref{dsarctwo}) is given to leading order by
\begin{equation}\label{laterdsdot}
\dot{R}_{dS, late}= - a^2 \left({L\, R_{dS, late}\over \pi M}\right)^{3/2}  
\end{equation} 
On the other hand, radial null rays in the KKdS black hole metric (\ref{kkmetric})  propagate according to $a\,\dot{\rho}_\pm = \pm U^{-3/2}$. In the late time limit when $z\ll 1$ for the deSitter horizon, one has $U\approx 1/z$ and this becomes
\begin{equation}\label{latenull}
\dot{\rho}_\pm =\pm a^2 \left({L\, \rho\over \pi M} \right)^{3/2}
\end{equation}
where the plus and minus signs correspond respectively to in and outgoing radial null rays.   Equation (\ref{laterdsdot}) is readily seen to have the same form as (\ref{latenull}) for ingoing null rays.  Hence an ingoing radial null ray is tangent to the deSitter apparent horizon in this late time limit.  Since  non-radial ingoing geodesics will have slower rates of progress in the radial direction, it follows that the deSitter apparent horizon in this regime traces out a future horizon, with timelike or null worldlines unable to cross in from outside the horizon.



Let us summarize the behavior of the deSitter apparent horizon over the different time ranges.
For early and intermediate times the tangent vectors to the approximate solutions we have found for the apparent horizon are outward propagating radial null vectors, while at late times the tangent  is an inward propagating null vector.
Since a number of approximations have been made in our analytical work,
we have also computed the tangent numerically throughout the evolution of the apparent horizon. The 
numerical work shows that the surface swept out by the apparent horizons is always timelike, asymptoting to an outgoing or an ingoing null surface 
in the limits of time going to minus or plus infinity respectively.
 
What does this behavior of the deSitter apparent horizon imply for the evolution of  deSitter event horizons?
First, recall the causal structure of the isolated
RNdS spacetime shown in figure (\ref{deflatingpatch}). The past and future
deSitter horizons intersect
at a bifurcation sphere which is at $r=\infty$ and $t=\infty$ in cosmological coordinates.
Since these are Killing horizons the cross sectional area is constant and $\theta =0$, so that  each spatial cross-section of the
Killing horizon is an apparent horizon. The KKdS black hole, on the other hand, is not 
static,  and the cross sectional area of an event horizon will generally change with time.   Whenever the horizon
area is changing, the expansion  $\theta$ will be non-zero. 
Hence, the $\theta =0$ apparent horizons we have studied coincide with the
event horizons only in such limits that the area of the deSitter apparent horizon is constant.
At early times,  the deSitter apparent horizon is a sphere with constant area given by (\ref{areasphcos}), while at late times it is a topological cylinder with asymptotically constant area given in (\ref{lateareaarc}) (with $a$ set to zero).   We expect that in these limits the family of deSitter apparent horizons we have found is close to the position of actual deSitter event horizons.  With reference to figure (\ref{deflatingpatch}), the family of deSitter apparent horizons appear to have turned the corner between the past and future deSitter event horizons, with the future deSitter horizon having been pulled in to finite coordinate radius.  A more precise understanding of this aspect of the causal structure of the KKdS spacetimes is left for future work.
  
%


\subsection{Cylindrical solutions}

It is also possible to find apparent horizons with cylindrical symmetry in the far field regime.  At intermediate times these cylindrical solutions match  those in the KKdS spacetime in section (\ref{kkdssec}), while at late times they asymptote to a constant comoving radius.   However, we do not find a limit in which the hypersurface swept out by the cylindrical horizons becomes null.  In the intermediate time range the surface is timelike, whereas in the late time regime it is spacelike.
%
Hence the cylindrical horizons do not appear to play a role in delineating the causal structure of the spacetime. Hence these simple cylindrical solutions provide an example that not all apparent horizons, even ones with a high degree of symmetry, coincide with an event horizon.  For completeness, we present the analysis of the cylindrical apparent horizons in the appendix.

\section{The Clayton effect - a singular  string}\label{string}%

We have found that, contrary to our initial intuition, the black hole apparent horizon of the KKdS black hole spacetime does not merge with itself around the compact direction as the universe collapses.  In our analytical work, we saw that the initial expansion of the black hole apparent horizon at early times reaches an endpoint, with a roughly spherical horizon of a finite limiting size, localized in the compact direction.  Nor, according to our  numerical work, does an apparent horizon of cylindrical topology subsequently formed outside this to complete a self-merger around the compact circle.  

All of our work has been in the limit that the parameter $\epsilon=MH_0^2\ll 1$.  In this same limit, a pair of black holes is known to merge  \cite{Kastor:1992nn,Brill:1993tm,Nakao:1994mm}.  If each has mass $M$, then although the individual black hole apparent horizons similarly stop growing after reaching a finite limiting size, a larger enveloping horizon subsequently forms and expands outward, eventually matching onto the event horizon of a single RNdS black hole of mass $2M$ in the far field, which represents the end state of the merger. 

However, in the KKdS black hole case, there is no well behaved final state for the black holes to merge into.
The natural guess for an end state of the  KKdS black hole horizon is the horizon of a charged black string.   In the present section, however,  we analyze the properties of the ``would be'' final state of the system, showing that it represents a naked singularity, rather than a black hole.  The reason for the failure of the KKdS black hole apparent horizon to self-merge can then be described in different, but essentially equivalent, ways.  We could say that the horizon does not merge with itself because there is no end state available with cylindrical topology.  Alternatively, we could say that the horizons stay separate in order to avoid evolution to a naked singularity.  This latter description, in which harmful mergers are prevented,  is what we have in mind by referring to this behavior the Clayton effect\footnote{Another example of the Clayton effect takes place in the case of two black holes such that the sum of the masses is greater than the maximum mass for a single black hole.  In this case  \cite{Nakao:1994mm} the individual black hole apparent horizons similarly fail to merge.}.

If one considers a continuous distribution of mass on a line, rather than the discrete set in (\ref{lattice}),  one obtains a uniform charged string. The metric  has the form (\ref{kkmetric}) with the harmonic metric function
\begin{equation}
U=1 + {\mu\over a^2\rho}\,,
\end{equation}
where we have defined the mass per unit length $\mu\equiv \pi\,M/L$. This is the same as the long distance limit of the Kaluza-Klein black hole metric.
For $H_0=0$ this is a static, charged string wrapping the compact direction.  It was shown in \cite{Myers:1986rx} that this spacetime has a naked singularity at $\rho=0$.  In fact, it was argued more generally, following a similar  discussion in the $D=4$ case \cite{Hartle:1972ya}, that any configuration with extended rather than point sources gives naked singularities.

We can see that this remains the case for $H_0\neq 0$  as well.  First, one finds that the square of the electromagnetic field strength diverges like 
$F^2\simeq -6/\mu\rho$, showing that the locus $\rho=0$ of the line source is singular.  Second, computing the expansion one finds
\begin{equation}
\theta= {1\over \rho^{2}\,a^3\,U^{3/2}}\left( 2\,a^2\rho +{\mu\over 2}\right) -3\,H_0
\end{equation}
which is easily seen to diverge near $\rho=0$ and approach a constant negative value $-3 \,H_0$ for large $\rho$.  A detailed examination of the expansion shows that it crosses zero only once at a deSitter-like apparent horizon.  Therefore, we expect that the singularity at $\rho=0$ is naked.
There is no cylindrical black hole horizon in the far field for the KKdS black hole apparent horizon to self-merge into\footnote{Note that an expanding deSitter apparent horizon does exist in the string spacetime.  One can check that the deSitter apparent horizon of the KKdS black hole does appropriately match onto this in the far field limit.}.

\subsection*{Acknowledgements}

This work   was supported in part by NSF grant PHY-0555304.



\renewcommand{\theequation}{A\arabic{equation}}
  \setcounter{equation}{0}  
  \section*{Appendix A: Cylindrical apparent horizons}  

Although the apparent horizons with cylindrical symmetry do not appear to play a role in the causal structure of the KKdS black hole spacetimes, as noted in section (\ref{kkdssec}), they may be of interest in the Kaluza-Klein reduction of the spacetime.  The analysis proceeds as follows.
We want to solve for cylinders with vanishing expansion (\ref{theta})  in the far field limit where $f\simeq \pi M/\rho L$ and $\rho/L\ll 1$.
In this case, let us assume that  $\rho=\rho_h$  is a constant and hence that $\rho '  (w)=\rho '' (w) =0$.   The equation $\theta=0$ then becomes
\begin{equation}\label{cyltheta}
a\,H_0\,\rho_h\,U^{1/2} =  {2\over 3} + {\rho_h\partial _\rho f\over 2\, a^2\,U} \  .
\end{equation}
%
Substituting in the far field limit for $f$ allows the equation to be written as
\begin{equation}\label{cylthetathree}
\frac{1+4\,z}{z^{1/2}\,(1+z)^{3/2}}= {6\pi\epsilon \over H_0\,a\,L}\,\,,
\end{equation}
where we have set $z =a^2\,\rho_h L/\pi M$. Since the function of $z$ on its left hand side is monotonic, equation (\ref{cylthetathree}) admits only one solution. 

If we look in a time regime such that  $6\pi\epsilon \ll H_0aL$, then there will be a solution to   (\ref{cylthetathree})  with $z\gg 1$  given approximately by  
%
\begin{equation}\label{earlycyl}
\rho_h \simeq {2\over 3\,a\,H_0}
\end{equation}
which matches the cylindrical horizon radius (\ref{cylindrical}) in the KKdS spacetime.  
Requiring that this radius be in the far field limit, imposes an upper bound on the scale factor as well, so that the solution (\ref{earlycyl}) is valid in the intermediate time range $6\pi\,\epsilon\ll H_0\,a\,L\ll 2/3$, which is consistent so long as the parameter $\epsilon$ is sufficiently small.

\begin{figure}
\centering
\subfigure
{
    \label{fig:sub:a}
    \includegraphics[width=7cm]{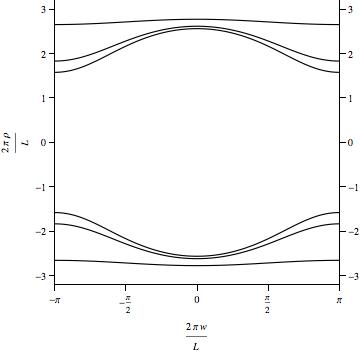}
}
\hspace{1cm}
\subfigure
{
    \label{fig:sub:b}
    \includegraphics[width=7cm]{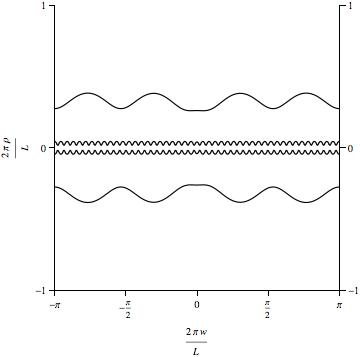}
}
\caption{\it 
Evolution of the cylindrical cosmological horizons for the choice of parameters $H=0.2\times(2\,\pi/L)$, $M=(L/2\,\pi)^2$. For this choice of parameters, $\epsilon=0.04$ and the condition $6\,\pi\,\epsilon< 2/3$ is marginally violated. Nevertheless the location of the horizons approximately follows the analytical estimate~(\ref{cylthetathree}). In the left panel, from the outermost to the innermost horizon the scale factor $a$ takes the values $1,\,0.1,\,0.01$. One can see that at late times the shape of the horizon becomes time-independent. In the right panel (note the different scale) the inner horizon corresponds to $a=100$, the outer to $a=10$. While the approximate location of the horizon follows eq.~(\ref{earlycyl}), its shape has a wavy form. This illustrates the fact that there are families of constant mean curvature surfaces which are solutions to $\theta =0$, but likely are not good approximations to the event horizon.}
\end{figure}

 
On the other hand when the scale factor is small, satisfying $H_0aL \ll 6 \pi\epsilon$, then the  comoving radius of the cylindrical apparent horizon in this late time limit goes to a constant value with
\begin{equation}\label{latecyl}
{\rho_h\over L}\simeq {1\over 36\,\pi \epsilon}
 \left[ 1+5 \left( {H_0aL \over  6\pi \epsilon} \right)^2 +{\cal {O}}\left(\left( {H_0aL \over  6\pi \epsilon}\right)^4\right)\right] 
\end{equation}
%
To see that these are apparent horizons of cosmological type, we restore the inequality in the  relation $\theta \geq 0$, which becomes $\frac{(1+4\,z)}{z^{1/2} (1+z )^{3/2}}\geq 6\pi \epsilon/H_0aL$. This implies that $\theta$ is positive for coordinates $\rho$ less than $\rho_h$ in both (\ref{earlycyl}) and (\ref{latecyl}).

In order to understand the causality properties of this horizon, it is convenient to compute the interval along its world volume, which is given by
\begin{equation}
ds_h^2=\frac{dt^2}{U^2}\,\left(-1+a^2\,U^3\,\left(\frac{d\rho_h}{dt}\right)^2\right)\,\,.
\end{equation}
In the intermediate time regime we obtain $ds_h^2=-5\,dt^2/9$, and the horizon is timelike as in the KKdS spacetime.  These are not null surfaces and light signals can cross in both directions.  At late times, on the other hand, one finds that $ds_h^2=16\,dt^2/9\,U^2$, which is spacelike.


\end{document}